\documentclass[aps,preprint]{revtex4}
\usepackage{graphicx}

\begin{document}

\title{Intermittent exploration on a scale-free network}

\author{A. Ramezanpour}
\email{aramezan@ictp.trieste.it}

 \affiliation{Institute for Advanced Studies in Basic Sciences, Zanjan 45195-1159, Iran}
 \affiliation{The Abdus Salam International Centre for Theoretical Physics, Strada Costiera 11, 34014 Trieste, Italy}

\date{\today}

\begin{abstract}
We study an intermittent random walk on a random network
of scale-free degree distribution. The walk is a combination of
simple random walks of duration $t_w$ and random long-range jumps.
While the time the walker needs to cover all the nodes increases
with $t_w$, the corresponding time for the edges displays a
non monotonic behavior with a minimum for some nontrivial value of
$t_w$. This is a heterogeneity-induced effect that is not observed
in homogeneous small-world networks. The optimal $t_w$ increases
with the degree of assortativity in the network. Depending on the
nature of degree correlations and the elapsed time the walker finds
an over/under-estimate of the degree distribution exponent.
\end{abstract}

 \maketitle

\section{Introduction}\label{1}
Random walks on regular and homogeneous structures have well been
investigated in the past studies \cite{h}. However, it has been
found that real networks such as World Wide Web and Internet, to
name a few, possess a random and heterogeneous structure
\cite{ab,n1}. Degree distribution in these networks usually exhibits
a power law behavior as $p(k)\propto k^{-\gamma}$, where usually
$\gamma$ lies between $2$ and $3$. Here $k$
denotes the node's degree (number of neighbors).

The network structure can significantly affect the behavior of a
random walker on it \cite{jsb,pa,alph,plh,t,nr,ch,bav}. For
instance, if a network has an effectively infinite dimension (as
happens for real networks), then the number of visited nodes by a
random walker may increase linearly with the number of time steps
\cite{pa,alph}. As another example, we know that it is easier to
find a target in some kinds of small-world networks \cite{k}. On the
other hand, for a fixed structure one may change the dynamics rules
to find better strategies to locate the target. Indeed, it has been
shown that intermittent walks could give a greater efficiency than
usual simple random walks in homogeneous structures
\cite{cbvm,svrl,blmv}. In this work we will introduce an
intermittent random walk and study its behavior on a random network
of scale-free degree distribution. We will see how the cover times
(to visit all the nodes or edges) change with the network structure
and the random walk dynamics.

Consider an explorer that starts its exploration from an arbitrary
page in WWW. The explorer usually goes through a shortcut to another
web page linked in the first one. She might do the same in the
new page or log off the Web. Usually, one does the exploration for a
limited time $t_w$ and the next time starts the exploration from another
page. This behavior of an explorer can be modeled by an
intermittent random walk on the WWW network. Here $t_w$ controls the random walk
intermittency. For very small $t_w$, random
long-range jumps dominate the walk, and if $t_w$ is very large one
recovers the usual random walk behavior.

Let the above intermittent explorer walk on a given network. We can
ask the following questions: How long does it take for the explorer
to cover all the nodes/edges in the network? There exist nodes of
very different degrees in a heterogeneous network.  How fast are
nodes of different degrees, or edges connecting nodes of given
degrees, visited by the above explorer? After a time the explorer
covers a portion of the original network. Looking at this visited
part we could obtain an estimate of the degree distribution. In the
case of scale-free networks, we ask how much the estimated exponent
of degree distribution is close to the original one. A similar
question has recently been asked for other quantities of a network
such as clustering and average degree in Ref. \cite{lfc}. But, in
that study the explorer is a random walker with no random long-range
jumps, i.e., $t_w=\infty$.

In the following we are going to address some of the above questions in a scale-free network. 
We will see that heterogeneity and degree
correlations between neighboring nodes have significant effects on
the behavior of the explorer. Interestingly, we find that there is an optimal
value of $t_w$ that minimizes the edge's cover time. This behavior originates
from the heterogeneous structure of the network.

The paper is organized in the following way: First we define more
exactly the model and networks that are studied in this work. Then
we discuss the results of numerical simulations. A summary of the
results and some concluding remarks are given in the conclusion.

\section{The model}\label{2}
Consider a given network and a random explorer that starts from an
arbitrary node at time step $t=0$. Then, for the next $t_w$ time
steps the explorer does a simple random walk. In a simple random
walk the walker just performs random local jumps; i.e. at
each time step the explorer selects, with equal probability, one of
the neighbors of the node visited at the previous time step.
At the end of this simple random walk, i.e. at time step $t=t_w$, the explorer
jumps instantaneously to a randomly selected node in the network and
again does a simple random walk for the next $t_w$ steps. This
process may continue for $t\equiv (w-1)t_w+l$ time steps. Here
$w=1,2,\ldots$ indexes the number of simple walks and $l=0, \ldots,
t_w$ counts the number of steps in the current simple walk. Notice
that long-range jumps occur instantaneously and $t$ gives the number
of local jumps along the edges of the network. In general one can
also assign a time $t_{l}$ to the long-range jumps. Here we are only
interested in the effects induced by the network structure and so we
will take $t_l=0$.

We call the above walk an intermittent random walk. It is clear that
one recovers the usual random walk by approaching $t_w$ to infinity.
On the other hand, if $t_w=0$ the walker just performs random
long-range jumps. In this case, by definition we have $t=0$ and no
edge of the network is visited at all.

At any time step $t$ of the walk we can define the fraction of
visited nodes and edges that are denoted by $\rho_n(t)$ and
$\rho_e(t)$, respectively.  We define the edge's cover time $t_e$ as the number of
necessary time steps to visit all edges. This time step is larger
than or equal to $t_n$, the time step at which all nodes have been
visited. Looking at the visited nodes and edges, the explorer finds
an estimated degree distribution of the network, denoted by
$p_k(t)$. The estimated exponent of the degree distribution,
$\gamma_e$, is extracted from the behavior of $p_k(t)$ for large
$k$'s.

We are going to study the above random walk on scale-free networks
of fixed degree distribution possessing different kinds of degree
correlations. We start with the network model introduced by
Barab\'{a}si and Albert (BA model) \cite{ba}; it is a growth model
where nodes of degree $m$ are successively added to the network.
Each edge of the new node is connected with probability
$\pi_i\equiv k_i/(\sum_j k_j)$ to a node of degree $k_i$ already present
in the network. This preferential attachment of the edges results in
a scale-free degree distribution, $p_k\propto k^{-3}$, for
sufficiently large number of nodes. Here we will consider networks of size $N=10^4$.

Generating good scale-free networks with small exponent $\gamma$ 
is not a trivial task especially when $N$ is not very large.
Indeed for $\gamma \le 3$ the second moment of degree distribution 
diverges in the thermodynamic limit. This in turn results to large fluctuations
in the tail of degree distribution for small $N$. It is why we selected
a scale-free network with the relatively large exponent $\gamma=3$. However, as we will see,
the main result of this study originates from the heterogeneous structure of the network.
Therefore we expect to find, qualitatively, similar results for smaller values of $\gamma$.

There is a little tendency for
nodes of dissimilar degree to be connected to each other in the above
network model. The correlation coefficient gives a measure of this
tendency and is defined as \cite{n2}
\begin{equation}
r\equiv\frac{\sum_{k,k'}kk'p_{k,k'}-(\sum_k
kq_k)^2}{\sum_{k}k^2q_{k}-(\sum_k kq_k)^2}.
\end{equation}
Here $p_{k,k'}=(1+\delta_{k,k'})E_{k,k'}/(2E)$ is the probability of
having an edge with end point nodes of degree $(k,k')$; $E_{k,k'}$
is the number of such edges in the network and $E$ is the total
number of edges.  The probability of finding a node of degree $k$ at
the end of an edge is $q_k=kp_k/\langle k \rangle$ where $\langle k\rangle$ denotes the average
degree. In uncorrelated networks $p_{k,k'}=q_kq_{k'}$ and so $r=0$.
For assortative and disassortative networks we have $0<r\le 1$ and
$-1<r<0$, respectively.

\begin{table}

\caption{Correlation coefficient of the networks studied in this
work. The networks size is $N=10^4$. The correlated and
uncorrelated networks have been obtained from BA model as
described in the text. The initial seed for growing BA model was a
pair of connected nodes and  $m=2$. Statistical errors are less
than $0.002$ and $Q=10^3 E$.}\label{tab.1}

\begin{center}

\begin{tabular}{|c|c|c|c|c|}

\hline
       & BA model & disassortative  & uncorrelated & assortative \\
\hline

 $r$    & $-0.045$ & $-0.120$        & $-0.019$     & $+0.156$ \\

\hline

\end{tabular}

\end{center}
\end{table}

To generate networks of different correlations we go through the
following instruction \cite{plh}: Starting from an arbitrary node,
say $x$, we randomly select  one of its neighbors $y$. Then the edge
between them is replaced by another one that connects $y$ to a
randomly selected node $z$, which is not already a neighbor of $y$.
We set $x=z$ and again go through the above process. Repeating the
process for a large number of times, $Q$, results in nearly
uncorrelated networks. To generate disassortative networks we
connect $y$ and $z$ only when $|k_z-k_y|>2$, otherwise neglect $z$
and select another node.  If we are to generate assortative networks
we connect $y$ and $z$ with probability
$[\it{min}(k_z,k_y)/\it{max}(k_z,k_y)]^{1.5}$. In Table \ref{tab.1}
we have given the correlation coefficients for networks that are
studied in this work.

\section{Discussion of the numerical results}\label{3}

First, let us see what happens for the intermittent random walker on
an uncorrelated random network of degree distribution $p_k$. For such a network,
the probability of finding the walker on a node of degree $k$ after
a random long-range jump is just $p_k$. Consider a node of degree
$k$ in the network. We define $r_k(t)$ as the the probability of finding the
walker on this node at time $t$. 
If $t_w=0$ then $r_k(t)= 1/N$. In this case the walker does not
differ between nodes of different degree and so all nodes are
visited  with the same rate. On the other hand, for $t_w=\infty$ the
walker has more chance to be found on a node of high degree. For
very large $t$ we have $r_k=k/(N\langle k \rangle)$ and for small $t$ we expect
$r_k$ to grow even faster than $k$. 
Now, high degree nodes are
visited more rapidly and low degree nodes are visited rarely.
Indeed, any deviation from $t_w=0$ decreases/increases the rate of
visiting low/high degree nodes. Usually a significant fraction of
the nodes have a low degree and so one expects the node's cover time
to be an increasing function of $t_w$.

Now consider an edge that connects two nodes of degree $k$ and $k'$
to each other. Again we define $r_{k,k'}(t)$ as the probability of visiting the edge at time step $t$.
As stated before the case $t_w=0$ is trivial because
all edges remain unvisited and so the edge's cover time is infinity.
However, when $t_w=1$ we have

\begin{equation}
r_{k,k'}(t)=\frac{1}{N}(\frac{1}{k}+\frac{1}{k'}).
\end{equation}

We see that the probability of visiting an edge with end point nodes
of high degree is very small. In the case of a scale-free network,
after a long-range jump the walker usually finds itself on a low
degree node. And if $t_w$ is very small the walker would not have
enough time to visit edges that connect high degree nodes. This
in turn results to a large cover time for the edges. On the other
hand, when $t_w=\infty$ 

\begin{equation}
r_{k,k'}(t)=r_k(t-1)\frac{1}{k}+r_{k'}(t-1)\frac{1}{k'}.
\end{equation}

If $r_k(t)$ grows faster than $k$ (we have numerically checked that it is indeed the case)
then the edges connecting low degree nodes have a very small 
chance to be visited by the walker. In this case the walker usually visits the
edges connecting high degree nodes. Therefore, the walker would need a
large time to cover the entire set of the edges.

Let us find an expression for $r_k(t)$ when $t_w=\infty$. The probability of
finding the walker on a node of degree $k$ is given by $q_k$. However, if we restrice ourselves 
to the set of visited nodes at time step $t$, then we can write 

\begin{equation}
Np_k r_k(t)\approx k \frac{N_k(t)}{N(t)\langle k \rangle(t)}.
\end{equation}

Here $N(t)$ is the total number of visited nodes and $N_k(t)$ denotes the number of visited nodes of degree $k$. The average degree $\langle k\rangle(t)$ is given by $\sum_k k N_k(t)/N(t)$. 
In a mean field approximation,  we have

\begin{equation}
N_k(t+1)=N_k(t)+q_k(1-\frac{N_k(t)}{Np_k}),
\end{equation}

where $1-\frac{N_k(t)}{Np_k}$ is the probability that the new visited node is visited for the first time. In this way we obtain 

\begin{equation}
N_k(t)=Np_k(1-e^{-\frac{k}{N\langle k\rangle} t}).
\end{equation}

And for $r_k(t)$ we find

\begin{equation}
r_k(t)\approx \frac{k}{N(t)\langle k \rangle(t)}(1-e^{-\frac{k}{N\langle k\rangle} t}).
\end{equation}

We see that even in a mean field approximation $r_k(t)$ increases faster than $k$ for any finite value of $t$.

The above arguments suggest that there may exist an optimal $t_w$
that minimizes the edge's cover time.  In the following we resort to
numerical simulations to see what happens for intermediate values of
$t_w$ and different network structures.

\begin{figure}
\includegraphics[width=15cm]{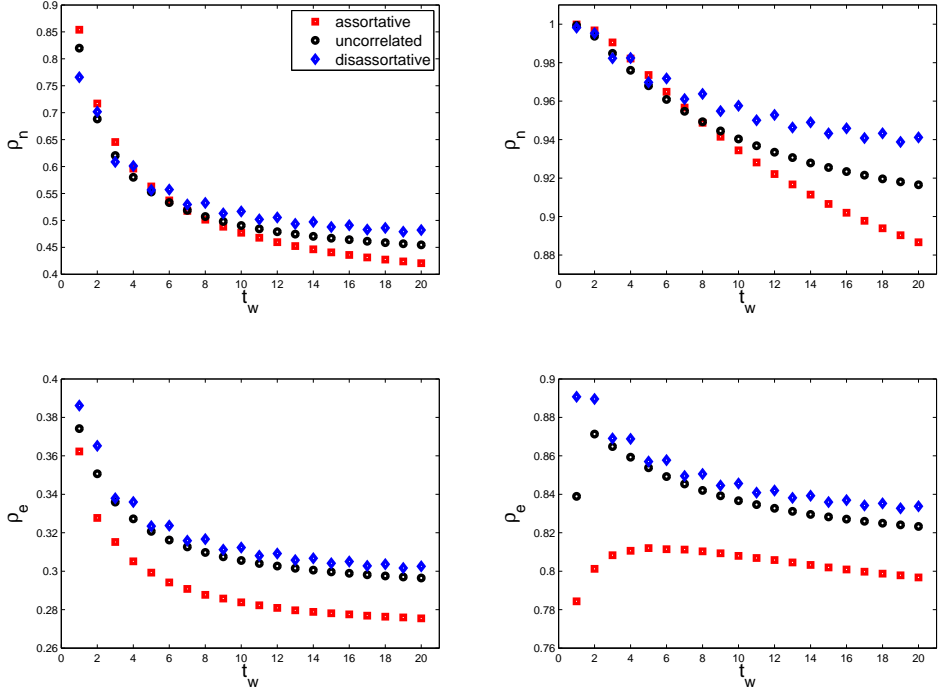}
\caption{The fraction of visited
nodes (top) and edges (bottom) vs $t_w$ at two times $\tau\equiv t/N=1$ (left)
and $\tau=5$ (right). In all the figures the parameters are the same as those
of table \ref{tab.1}.}\label{f1}
\end{figure}

First we study how $\rho_n$ and $\rho_e$ behave with $t_w$ for some
fixed value of $t$. Figure \ref{f1} shows the results of numerical
simulations for these quantities. We see that $\rho_n$ always
decreases with $t_w$, and except for small $t_w$'s, the explorer
visits a smaller number of nodes in assortative networks. In fact,
for a large $t_w$ the walk is mostly limited to the core of high
degree nodes. So the explorer would miss the other nodes which make
a major part of the network.

As Fig. \ref{f1} shows, the situation is more interesting for the
fraction of visited edges. While at small times $\rho_e$ decreases
with $t_w$, for larger times it develops a maximum at some
intermediate value of $t_w$. This phenomenon is clearer in the
assortative networks.

\begin{figure}
\includegraphics[width=15cm]{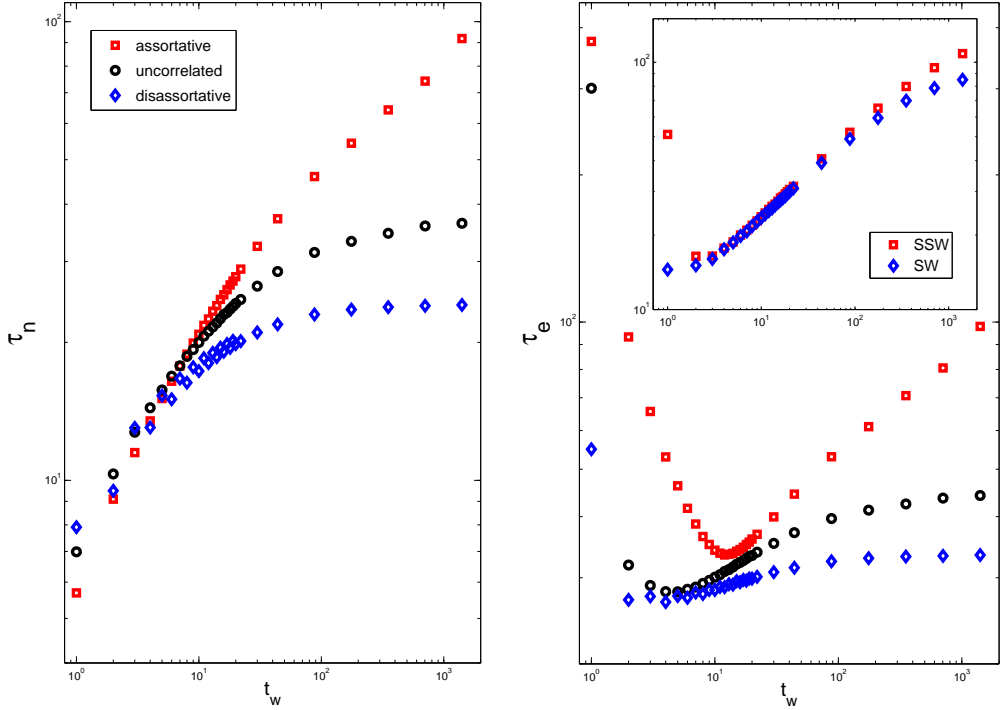}
\caption{Cover times $\tau_n$ (left)
and $\tau_e$ (right) vs $t_w$.  Inset (left): $SW$ and $SSW$ refer
to small-world and smallest small-world networks, respectively. The
number of nodes and shortcuts are $N=10^4$ and $M=10^3$
respectively. Statistical errors are about a few percent.}\label{f2}
\end{figure}

The presence of a maximum in $\rho_e$ brings the hope to have a
minimum $t_e$ for some $t_w$. Figure \ref{f2} shows the cover times
$\tau_n\equiv t_n/N$ and $\tau_e\equiv t_e/N$ versus $t_w$. The
results for different kinds of degree correlations have been
compared in this figure. As expected, in all cases $\tau_n$
increases monotonically with $t_w$. There are points at which
increasing $t_w$ does not result to any change in $\tau_n$. In this
case $t_w$ is large enough for the explorer to have the chance to
visit different parts of the network. Moreover, as Fig. \ref{f2}
shows, the assortative networks display this saturation phenomenon
at a larger $t_w$, and again this is due to the presence of a core
of high degree nodes in those networks. In a disassortative network
a community of nodes is usually formed by some low degree nodes
gathering around a high degree one. These communities are usually
connected to each other by a few edges. A characteristic time here
is the time the random walker needs for escaping a typical community.
If $t_w$ is greater than this time, the walker behaves as if it had
more random jumps during its walk.

In Fig. \ref{f2} we also observe the behavior of $\tau_e$ with
$t_w$. As expected, we find that in a scale-free network there is an
optimal $t_w$ for covering the edges, and hence the whole network.
This phenomenon is more prominent in the assortative networks. When
$t_w$ is very small the explorer mostly visits the edges emanating
from the low degree nodes which have more chance to be selected as
the starting points of the simple random walks. On the other hand,
when $t_w$ is very large, most of the time is spent on already
visited edges in the core of high degree nodes where the explorer is
found most of the time.

In Fig. \ref{f2} we also see that the optimal $t_w$ is
greater for the assortative networks. In fact, in disassortative
networks the explorer reaches a high degree node in a few steps
after starting a simple random walk. The time to reach a high degree node
would be larger in assortative networks where low degree nodes are
more likely to be connected to each other.

To show that the observed non monotonic behavior of $\tau_e$ is
indeed a heterogeneity-induced phenomenon, we obtained $\tau_e$
for two other networks: the small-world network \cite{nmw} and the
smallest small-world network \cite{dm}. The former network, which
is a homogeneous one, is constructed by adding randomly $M$
shortcuts to a chain of $N$ nodes. The latter network, which is
not homogeneous, is constructed by adding $M$ shortcuts, but this
time emanating from a single node on the chain and distributed
randomly throughout the network. As the inset in Fig. \ref{f2}
shows, in the small-world network $\tau_e$ increases monotonically
with $t_w$ whereas it exhibits a minimum in the smallest
small-world network.

\begin{figure}
\includegraphics[width=15cm]{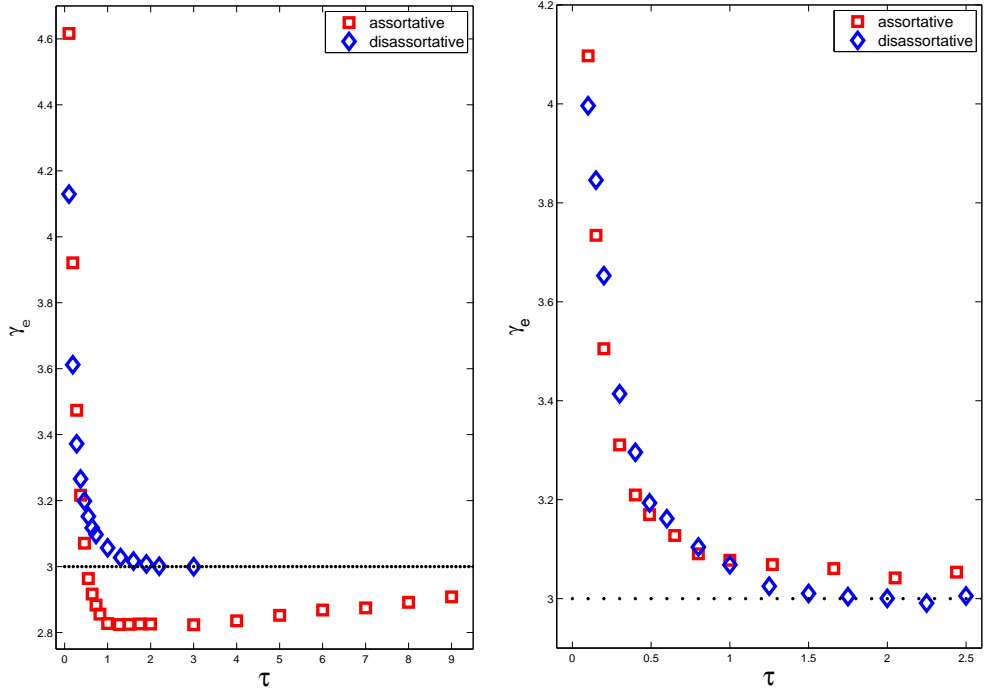}
\caption{Estimated $\gamma$ vs time
for $t_w=5$ (left) and $t_w=\infty$ (right).  Statistical errors are
of order $0.01$.}\label{f3}
\end{figure}

Finally we look at the estimated exponent of degree distribution by
the explorer. At a given time $t$ we construct a network of visited
nodes and edges. The estimated degree distribution $p_k(t)$ is
obtained from the constructed network. Then we fit a power law
function $k^{-\gamma}$ to the tail of this distribution. We use only
the data in the range $[k_{max}/4,k_{max}]$ where $k_{max}$ is the
maximum degree that appears in $p_k(t)$. Then the estimated $\gamma$
is obtained by a maximum likelihood procedure \cite{n3}. In Fig.
\ref{f3} we plot the estimated exponent versus $\tau\equiv t/N$ for
$t_w=5$ and $t_w=\infty$. There, we have compared the results for the
assortative and disassortative networks with $\gamma_0=3.003\pm
0.004$. This exponent has been obtained for the original network
with the same procedure described above.

Figure \ref{f3} shows that at short times the estimated exponent is
larger than the original one in both the assortative and disassortative networks.
For $t_w=5$ and small $t$, the fraction of visited edges is small and most of them are incident
on low degree nodes. The probability of having a high degree node
is very small and so the visited network seems more homogeneous
than the original one. In this limit the estimated
$\gamma$ is much larger in the assortative networks. This stems
from the fact that the assortative networks are locally more
homogeneous than the disassortative ones. As Fig. \ref{f3} shows, the
scenario changes in long times. We have an underestimate of the
exponent in the assortative networks and still an overestimate in
the disassortative ones. For large times the new visited edges mostly
connect high degree nodes in the assortative networks.
Consequently, the fraction of high degree nodes in the visited network rapidly
increases. Notice that in the disassortative
networks the new visited edges usually connect
dissimilar nodes and so we have a uniform approach to the real
structure of the network.

In Fig. \ref{f3} we see that for $t_w=\infty$ 
the estimated exponent is always greater than the original
one. Here, there is no significant difference between the values of
$\gamma_e$ in the assortative and disassortative networks. Comparing
the two cases $t_w=5$ and $t_w=\infty$, we see that in the
disassortative networks the estimated exponent approaches the real
value with nearly the same rate. On an assortative network the
explorer finds a better estimate of $\gamma$ if it has a simple
random walk with no long-range jumps.

\section{Conclusion}\label{5}
In summary, we showed that there is an optimal $t_w$ that
minimizes the cover time $t_e$ in a scale-free network. The
optimal $t_w$ increases with the degree of assortativity in the
network. This non monotonic behavior of $t_e$ with $t_w$ originates
from the heterogeneous structure of the network. Therefore,  we expect that
increasing (decreasing) $\gamma$ will weaken (strengthen) the
observed effect. Depending on the nature of degree correlations in the
network we could obtain an overestimate or underestimate of the
exponent $\gamma$.

The findings might be useful in devising good strategies to cover
a heterogeneous network and find a good estimate of the network's structure. It
will be interesting to study more realistic cases where $t_w$
obeys a given distribution.

We have numerically checked that the non monotonic behavior of $t_e$
is also observed when $t_w$ follows an exponential distribution. The
above results are also qualitatively true for other values of $m$,
or the average degree. Moreover, the optimal $t_w$ is not very
sensitive to the size of network, $N$. For instance, if we increase
$N$ form $10^3$ to $10^4$, the optimal $t_w$ decreases nearly by
$2$. In this study we considered the case that long-rang jumps occur
instantaneously, i.e. $t_l=0$. The way that the covering times
behave with $t_w$ depends also on the magnitude of this quantity.
Clearly, for very large $t_l$, both $t_n$ and $t_e$ will decrease
with $t_w$.

\acknowledgments I would like to thank M. R. Kolahchi and B.
Farnudi for reading the previous version of the manuscript.
I am also grateful to M. Marsili for useful comments and discussions.

\end{document}